\documentclass[lettersize,journal]{IEEEtran}
\usepackage{amssymb,amsmath,amsthm,amsfonts}
\usepackage{algorithmic}
\usepackage[linesnumbered,ruled, boxed, lined]{algorithm2e}
\usepackage{array}
\usepackage[]{acronym}
\usepackage[caption=false,font=normalsize,labelfont=sf,textfont=sf]{subfig}
\usepackage{textcomp}
\usepackage{stfloats}
\usepackage{url}
\usepackage{verbatim}
\usepackage{graphicx}
\usepackage{cite}
\usepackage{tikz}
\DeclareMathOperator*{\argmax}{argmax}
\DeclareMathOperator*{\argmin}{argmin}

\usepackage{tikz}
\usetikzlibrary{shapes.geometric, arrows}

\tikzstyle{startstop} = [rectangle, rounded corners, minimum width=3cm, minimum height=1cm,text centered, text width=3cm, draw=black, fill=magenta!30]
\tikzstyle{io} = [trapezium, trapezium left angle=70, trapezium right angle=110, minimum width=1cm, minimum height=1cm, text centered, draw=black, fill=blue!30]

\tikzstyle{output} = [trapezium, trapezium left angle=70, trapezium right angle=110, minimum width=1cm, minimum height=1cm, text centered, draw=black, fill=purple!30 ]

\tikzstyle{process} = [rectangle, minimum width=3cm, minimum height=1cm, text centered,text width = 4cm, draw=black, fill=cyan!8]
\tikzstyle{decision} = [diamond, aspect=2,  minimum width=3cm, minimum height=0.5cm, text centered, draw=black, fill=green!10]
\tikzstyle{arrow} = [thick,->,>=stealth]
\tikzstyle{external} = [circle,  radius=2cm,  draw=black, fill=blue!85, text=white]
\tikzstyle{externalIN} = [circle,  radius=2cm,  draw=black, fill=green!85, text=black]
\tikzstyle{ioarrow} = [thick,->>,>=stealth, draw=blue]
\tikzstyle{arrowIN} = [thick,->>,>=stealth, draw=green]

\begin{document}
\acrodef{AWGN}{additive white Gaussian noise}
\acrodef{PLS}{physical layer security}
\acrodef{UCB}{upper-confidence bound}
\acrodef{RL}{reinforcement learning}
\acrodef{MAB}{multi-armed bandit}
\acrodef{WSC}{weighted secrecy coverage}
\acrodef{SINR}{signal-to-interference-plus-noise ratio}
\acrodef{PSO}{particle swarm optimization}
\acrodef{IoT}{Internet of Things}
\acrodef{UAV}{unmanned aerial vehicle}
\acrodef{ABS}{aerial base station}
\acrodef{FDMA}{frequency division multiple access}
\acrodef{QoS}{quality of service}
\acrodef{WCS}{worst-case scenario}
\acrodef{LLL}{log-linear learning}
\acrodef{SLLL}{synchronous log-linear learning}
\acrodef{BLLL}{binary log-linear learning}
\acrodef{SBR}{smooth best response}
\acrodef{SOP}{secrecy outage probability}
\acrodef{LoS}{line of sight}
\acrodef{NLoS}{non-line of sight}
\acrodef{JC}{jamming coverage}
\acrodef{JE}{jamming efficiency}
\acrodef{A2G}{air-to-ground}
\acrodef{SNR}{signal-to-noise ratio}
\acrodef{PLB}{positioning learning block}
\acrodef{mmWave}{millimeter wave}
\acrodef{NOMA}{non-orthogonal multiple access}
\acrodef{5G}{the fifth generation of wireless networks}
\acrodef{6G}{the sixth generation of wireless networks}
\acrodef{MEC}{multi-access edge computing}
\acrodef{ZF}{zero-forcing}
\acrodef{A}{Alice}
\acrodef{B}{Bob}
\acrodef{E}{Eve}
\acrodef{KKT}{Karush-Kuhn-Tucker}
\acrodef{ACK}{acknowledgment signal}

\title{Strategic Deployment of Swarm of UAVs for Secure IoT Networks}


\author{X. A. Flores Cabezas,~\IEEEmembership{Student Member,~IEEE,}
and D.~P.~Moya~Osorio,~\IEEEmembership{Senior Member,~IEEE}\vspace{-2em} 
\thanks{X. A. Flores Cabezas and Diana P. Moya~Osorio are with the Centre for Wireless Communications (CWC), University of Oulu, Finland (e-mails: \{xavier.florescabezas;diana.moyaosorio\}@oulu.fi).}
\thanks{This research has been supported by the Academy of Finland, 6G Flagship program under Grant 346208 and project FAITH under Grant 334280.}
}



\maketitle

\begin{abstract}
Security provisioning for low-complex and constrained devices in the Internet of Things (IoT) is exacerbating the concerns for the design of future wireless networks. To unveil the full potential of the sixth generation (6G), it is becoming even more evident that security measurements should be considered at all layers of the network. This work aims to contribute in this direction by investigating the employment of unmanned aerial vehicles (UAVs) for providing secure transmissions in ground IoT networks. Toward this purpose, it is considered that a set of UAVs acting as aerial base stations provide secure connectivity between the network and multiple ground nodes. Then, the association of IoT nodes, the 3D positioning of the UAVs and the power allocation of the UAVs are obtained by leveraging game theoretic and convex optimization-based tools with the goal of improving the secrecy of the system. It is shown that the proposed framework obtains better and more efficient secrecy performance over an IoT network than state-of-the-art greedy algorithms for positioning and association.
\end{abstract}

\begin{IEEEkeywords}
3D position control, IoT, node association, physical layer security, unmanned aerial vehicle.
\end{IEEEkeywords}

\section{Introduction}
\Ac{5G} is envisioned to bring upon ubiquitous connectivity. Looking forward, beyond \ac{5G}, great advancements have been envisioned for \ac{6G}, which promises ubiquitous intelligence~\cite{art:Porambage2021}. Toward that, many low-complexity wireless devices would be part of populated decentralized networks, where absolutely everything is connected in massive deployments of \ac{IoT} networks, with applications in very different sectors, namely, industry, defense, healthcare, intelligent transportation systems, to name a few~\cite{9097898}.

In such dense, heterogeneous networks, very sensitive information is transmitted over a shared medium, thus security and privacy issues become critical, and they cannot be handled independently of other parameters, i.e. energy consumption or latency~\cite{art:Porambage2021}. While traditional cryptographic approaches have developed to be trustable solutions for preserving security in communications, the limitations and constraints of \ac{IoT} devices and sensors, and the advancements in quantum computing render these approaches unfeasible or unreliable~\cite{9097898}. 
On the other hand, \ac{PLS} techniques, that explore the inherent properties of the noisy and random wireless channels to provide security to communications, has emerged as a promising and attractive security solution. Some well-known \ac{PLS} techniques include artificial noise injection through friendly jamming, spatial diversity, beamforming design and relaying~\cite{art:Porambage2021, art:Sun2019,9681822}. These techniques aim at designing the physical layer to provide an advantage of the legitimate link over the eavesdropping link with no assumption on the computing power of the attacker, thus providing information-theoretic security guarantees. 

From other perspective, it is recognized that \acp{UAV} will play an important role in IoT applications, specially to provide connectivity in remote areas, disaster zones, and harsh environments~\cite{9424181,uavenviron}. Thanks to their flexible deployment, capability of providing strong \ac{LoS} connectivity and, ease of maneuverability, \acp{UAV} open a new range of novel opportunities for wireless networks, but at the same time, novel threat vectors should be also considered~\cite{9681822}. Noting this advantageous properties, \acp{UAV} can also be exploited for the design of \ac{PLS} techniques to safeguard \ac{UAV}-assisted communications. For instance, the challenges and opportunities for preventing passive and active attacks in wireless networks have been recently discussed in~\cite{art:Sun2019}. 

Particularly, the introduction of \ac{UAV} nodes acting as friendly jammers in order to improve the secrecy performance of wireless networks have recently risen special attention~\cite{art:Zhou2018}. All in all, the integration of \acp{UAV} into the provisioning of security through \ac{PLS} techniques provides novel opportunities for safeguarding \ac{6G} networks. Importantly, the use of learning methods would allow the \acp{UAV} not only to remain autonomous, but also to adapt to the complexity of \ac{PLS} security provisioning under dynamic channels and complex \ac{IoT} scenarios, which is the main focus of this work.

\subsection{Related Work}

{\color{black}
Recently, the flexibility of UAVs have rised attention for secure transmissions in wireless networks~\cite{art:Wei_SecureUAV,art:Yoo_SecureUAVDRL,art:Chen_UAVSecureOffloading,art:Lu_SecureUAV,art:Salem_UAVMalicious,art:Dong_UAVDRLmmWave,art:Illi_PLSDualHopUAV}. In particular, UAVs }have been employed as friendly jammers to assist a legitimate transmission by introducing artificial noise in order to prevent leakage of information to possible eavesdroppers in the network~\cite{art:Zhou2018,UAV_Jammer_Li,UAV_MEC_Zhou,art:Pang2021,art:Kim2021,Vilela_Metrics,Flores_Euncn2021,Flores_Pimrc2021,Flores_Eurasip,art:Flores_TVT}. In~\cite{art:Zhou2018}, the optimal three-dimensional (3D) deployment and jamming power of a UAV-based jammer are investigated to improve the secrecy performance of a wireless network in terms of the outage probability and the intercept probability, by defining area-based metrics that ensure a given intercept probability threshold within a certain area. In~\cite{UAV_Jammer_Li}, a UAV friendly jammer scheme is introduced to enhance the secrecy rate of a wireless system, where the problem of trajectory optimization is investigated. {\color{black} In~\cite{UAV_MEC_Zhou}, a joint jamming scheme between the legitimate UAVs serving as \ac{MEC} servers and the ground nodes is proposed to safeguard the legitimate transmission against malicious UAVs. 
Therein, the minimum secrecy capacity among system users is maximized by jointly optimizing the position, jamming power, and the computing capacity of the legitimate UAV, as well as the offloading rate of the users to the UAV, the transmit power of the users, and the offloading user association. Therein, it was demonstrated that the max-min secrecy capacity is improved over the benchmarks, specially for low offloading requirements, while existing a trade-off between security and latency.} 
In~\cite{art:Pang2021}, the \ac{SOP} of a UAV-based \ac{mmWave} relay network in the presence of multiple eavesdroppers is investigated, where the scenarios with and without cooperative jamming were contrasted. 
In~\cite{art:Kim2021}, the existence of an optimal UAV jammer location on a network with multiple eavesdroppers was proven, and the impact of the density of eavesdroppers, the transmission power of the UAV jammer, and the density of UAV jammers on the optimal location was investigated. {\color{black}In~\cite{Vilela_Metrics}, two area-based secrecy metrics, the \ac{JC} and the \ac{JE}, were proposed to evaluate the impact of jamming for secure wireless communications based on the \ac{SOP} over an area, without knowledge of the position of the eavesdropper. Later, in~\cite{Flores_Euncn2021}, this idea was extended by introducing a hybrid secrecy metric, the so-called \ac{WSC}, that considers both coverage and efficiency of friendly jamming, simultaneously, in the context of UAV-based friendly jamming. Therein, the positioning of the UAV jammers to maximize the \ac{WSC} is tackled. Further, in~\cite{Flores_Pimrc2021}, a null-space precoding scheme is employed to eliminate the interference at the legitimate receiver. Under that scheme, a better performance was obtained in terms of the \ac{WSC}. Further, in~\cite{Flores_Eurasip} and \cite{art:Flores_TVT}, the previous scenario was extended to include the 3D movement of the UAVs and the movement of the legitimate ground user, respectively. These works consider the formulation of the problem of adaptive position control of the UAVs as a multi-armed bandit, and the results presented significant improvements of the secrecy of the system in terms of WSC. In~\cite{art:Li_NOMAUAV}, it is considered a system where a UAV is serving a group of ground users via \ac{NOMA}, while sending artificial noise to disrupt a passive eavesdropper in the system. The total jamming power and the rate at each user are maximized by optimizing the UAV trajectory, the power allocation, and the user scheduling. Such scheme was proven to outperform orthogonal multiple access schemes as well as non-jamming schemes in terms of the system sum-rate and of the eavesdropper data-rate. }

{\color{black}
In recent years, the use of machine learning techniques has been increasingly considered to optimize the deployment of UAVs in wireless networks~\cite{UAV_Federated,DeepRL_Zhao,GT_RL_Fragkos,DeepQ_Li,HammoutiGreedyAlgsUAV}. For instance, a novel federated learning-based framework for the distributed joint power allocation and scheduling of swarm of UAVs was proposed in~\cite{UAV_Federated}. The proposed framework significantly improves the convergence time of two baseline methods, namely optimized power-randomized scheduling and randomized power-optimized scheduling. In \cite{DeepRL_Zhao}, an actor-critic deep \ac{RL} approach is proposed to find the optimal trajectory design and power allocation in UAV-assisted cellular networks, which achieves better network performance in terms of the average sum-rate of the system. In \cite{GT_RL_Fragkos}, game theory and \ac{RL} are used to enhance the data offloading from UAVs to MEC servers in an \ac{IoT} scenario. Therein, it was proven that the proposed methods converge to a Nash equilibrium of average offloaded data, whereas the \ac{RL} approach ensures the convergence without exchange of information between UAVs. In \cite{DeepQ_Li}, a deep Q-Learning-based scheduling approach is used to minimize the packet loss of \ac{IoT} nodes in UAV-assisted wireless powered-\ac{IoT} networks. The deep Q-Learning algorithm performs \ac{IoT} node and modulation scheme selection for \ac{IoT} nodes that wish to send information and wirelessly receive power  from the UAVs. It was shown that the deep Q-Learning approach obtains much lesser packet loss than greedy or random scheduling approaches. In~\cite{HammoutiGreedyAlgsUAV}, the binary log-learning (BLLL) and greedy algorithms are proposed to maximize the total sum rate of the users throughout the network by optimizing the user-UAV association and UAV position control in a UAV-assisted network. Therein, it was shown that greedy algorithms for UAV position control and user-UAV association are sub-optimal and obtain a lower sum-rate than BLLL. However, the convergence of BLLL present an exponential time, thus the greedy algorithms are preferable in this aspect.
}

Also, a deep Q-Network-based power allocation strategy was proposed in~\cite{Xiao_DQN_UAV}, to improve the secrecy rate of a legitimate communication between a UAV and a mobile user in the presence of a malicious mobile user and UAV. Therein, it is assumed that the attackers can choose between eavesdropping, spoofing and jamming attacks, and the results proved to overcome benchmarks based on Q-Learning and a win or learn faster-policy hill climbing (WoLF-PHC) approach. More recently, the optimization of the sum secrecy rate of a system with a single \ac{UAV} acting as an \ac{ABS}, that serves a group of ground nodes in the presence of UAVs acting as adaptive eavesdroppers or jammers, was proposed in~\cite{Liu_GT_UAV}. Therein, a Stackelberg game was formulated considering two strategies, the ABS positioning to increase the sum secrecy rate of the system as the leader, and the cooperative attack of the adaptive eavesdroppers as the follower. Then, a spatial adaptive play learning algorithm is utilized to reach the equilibrium, which is shown to obtain a better sum secrecy rate than a random or ring deployment of the ABS. 

\subsection{Main Contributions}

{\color{black}
To contribute to the state-of-the-art, this work considers the association, power allocation, and position control of UAVs serving as \acp{ABS} to a set of ground IoT nodes through frequency division multiple access (FDMA), by focusing on the secrecy performance of the system. Different from the approach in \cite{HammoutiGreedyAlgsUAV}, in this work the sum secrecy rate of the network is considered as the utility function, and the power allocation per node is also investigated. Moreover, different from the works in~\cite{UAV_MEC_Zhou,Xiao_DQN_UAV,Liu_GT_UAV}, inactive nodes in the system are treated as potential eavesdroppers, thus presenting a relatively high density of eavesdroppers in the system. For the user-UAV association and UAV positioning, the synchronous log-linear learning (SLLL) formulation is considered, which is a synchronous algorithm that offers faster convergence.
}
All in all, the main contributions of this paper are three-fold:
 \begin{enumerate}
     \item A three-stage block-coordinate ascend (BCA) framework is proposed where node association, UAV 3D position control, and power allocation are the blocks that are optimized iteratively by considering the other blocks fixed in order to increase the sum secrecy rate and number of nodes with positive secrecy in the proposed network.
     \item Game-theoretic algorithms are proposed for node association and UAV position control to improve the secrecy capacity of the system.
     \item A convex optimization-based power allocation technique is developed to increase the minimum secrecy rate of IoT nodes that can achieve secrecy, while guaranteeing a level of service to all IoT nodes.
 \end{enumerate}

\section{System Model}

\begin{figure}[bt]
    \centering
    \includegraphics[width=0.95\linewidth]{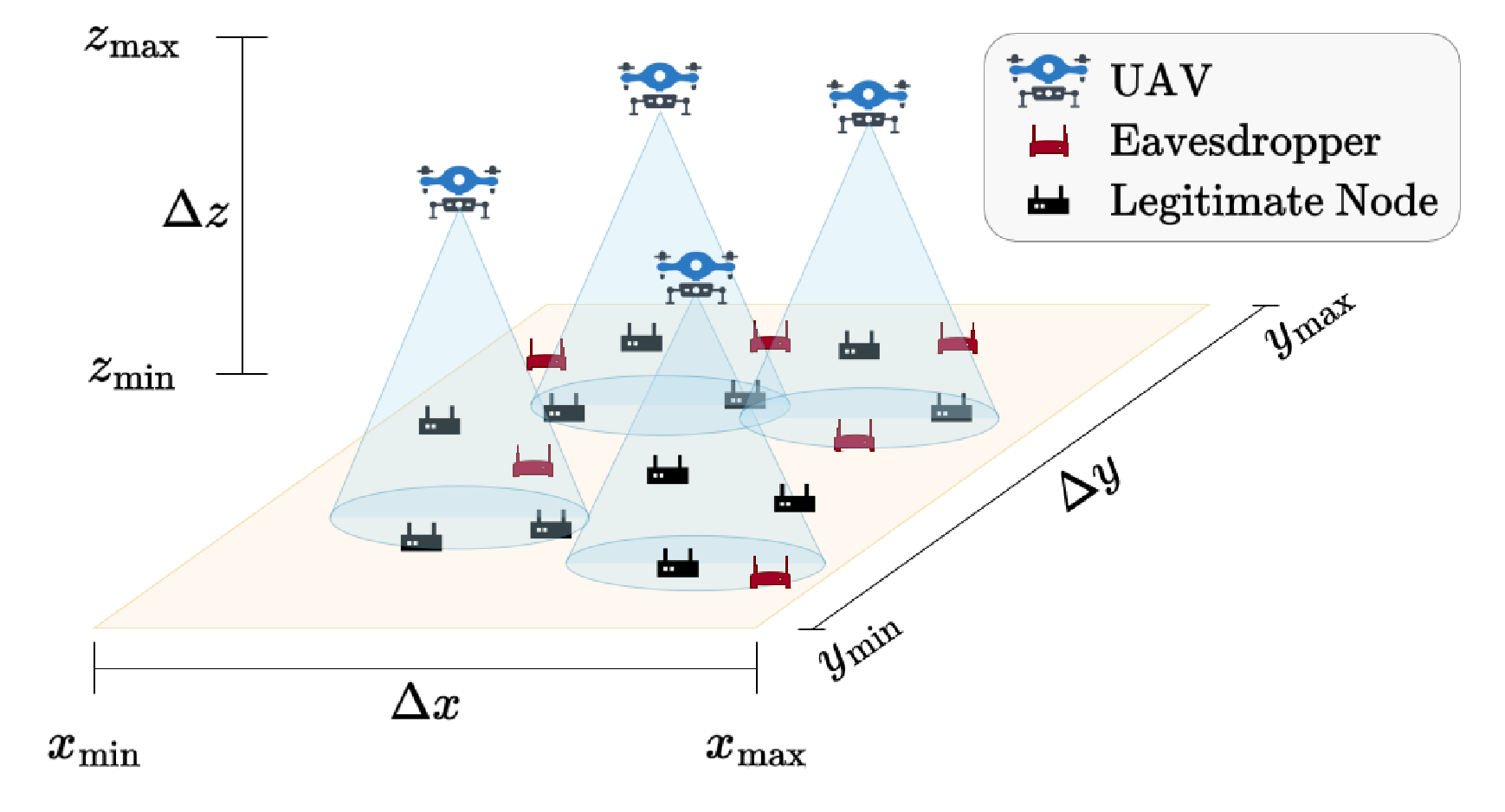}
    \caption{System model.}
    \label{fig:sysModel}
\end{figure}

Consider the system illustrated in Fig.~\ref{fig:sysModel}, which consists of a set of $N$ \ac{IoT} devices that are distributed following an uniform binomial point process over a rectangular region of dimensions $\Delta x = x_{\mathrm{max}} - x_{\mathrm{min}}$ and $\Delta y = y_{\mathrm{max}} - y_{\mathrm{min}}$, with the bi-dimensional position of the $n$th-\ac{IoT} device (that can be a legitimate node or eavesdropper) denoted by $\mathbf{x}_n = (x_n,y_n)$. To provide connectivity to the \ac{IoT} devices, a swarm of $M$ single-antenna \ac{UAV}s, acting as \ac{ABS}s is deployed over the region of interest. These \ac{UAV}s can move in three dimensions over the rectangular region, within a altitude range $\Delta z = z_{\mathrm{max}} - z_{\mathrm{min}}$. In this system, it is considered that, for a certain transmission process, only a fraction of \ac{IoT} devices (randomly and independently selected according to a Bernoulli distribution of parameter $q$) are set on receiving mode (legitimate nodes)
, while the rest are overhearing the channel, thus being considered as potential eavesdroppers.

In this system, downlink transmissions from the UAVs to the \ac{IoT} devices are based in \ac{FDMA}. Assuming that all UAVs have the same limited amount of bandwidth $BW$, each one divides its total bandwidth into $C$ ortogonal sub-channels of bandwidth $B=BW/C$. Additionally, let $\mathcal{N}$ be the set of ground nodes, while $\mathcal{L}$ and $\mathcal{E}$ are the sets of legitimate nodes and eavesdroppers, such that $|\mathcal{N}| = N$, $|\mathcal{L}| = L$ and $|\mathcal{E}| = E$, respectively. Additionally, $\mathcal{M}$ is the set of \ac{UAV}s, such that $|\mathcal{M}|=M$ and $\mathcal{C}$ is the set of sub-channels available at each \ac{UAV}, with $|\mathcal{C}| = C$. For simplicity purposes,  the described sets are treated as their respective sets of indices as well. 

Accordingly, each \ac{UAV} can associate with up to $C$ ground nodes, with the power allocated by \ac{UAV} $m\in\mathcal{M}$ to the sub-channel $c\in\mathcal{C}$ denoted as $p_m^c$, and the total power budget at each \ac{UAV} is $P$. Then, the power allocation vector at UAV $m$ is given by $\mathbf{p}_m = (p_m^1,...,p_m^{C})^T$ and the power allocation matrix of the whole system is given by $\mathbf{P}\in\mathbb{R}^{C\times M}$ with $\mathbf{P} = [\mathbf{p}_1, ..., \mathbf{p}_M]$. 
Let $\mathbf{A}\in \mathbb{R}^{L\times M\times C}$ be the association array with elements $a_{l,m,c}\in \{0,1\}$, where $a_{l,m,c}=1$ if node $l$ is associated to UAV $m$ through sub-channel $c$, and $0$ otherwise. Given that, at any time, a certain sub-channel is either available or assigned to a single node, and that all legitimate nodes are associated to a single sub-channel, it holds that
\begin{align}
    \label{eq:optProbconst1_1}\sum\limits_{l\in\mathcal{L}}a_{l,m,c} &\leq 1 \;\;\;\; \forall m\in\mathcal{M}, \forall c\in \mathcal{C},\\
    \label{eq:optProbconst2_1}\sum\limits_{m\in\mathcal{M}}\sum\limits_{c\in\mathcal{C}}a_{l,m,c} &\leq 1 \;\;\;\; \forall l\in\mathcal{L}.
\end{align}

The \ac{A2G} channel between UAV $m$, at altitude $z_m$, and a ground node $l$ is modeled as in~\cite{art:Zhou2018}, with $P_{\textrm{LoS}}$ and $P_{\textrm{NLoS}}$ probabilities of LoS and NLoS connection being, respectively, given by~\cite{art:Zhou2018}
\begin{align}
    P_{\textrm{LoS}} &= \frac{1}{1 + \psi \exp \left( -\omega\left[ \frac{180}{\pi}\tan^{-1}\left(\frac{z_m}{r_{m,l}}\right)-\psi \right] \right)}
\end{align}
and $P_{\textrm{NLoS}} = 1 - P_{\textrm{LoS}}$, with $\psi$ and $\omega$ being environmental constants~\cite{Dao_PropParameters,Hourani_PropParameters2}, and $r_{m,l}$ is the distance from node $l$ and the projection on the ground of UAV $m$. Then, the average pathloss of the links is given by
\begin{equation}
    L_{m,l} = \left( z_m^2 + r_{m,l}^2 \right)^{\frac{\alpha_J}{2}} \left(P_{\mathrm{LoS}}\eta_{\mathrm{LoS}} + P_{\mathrm{NLoS}}\eta_{\mathrm{NLoS}} \right),
\end{equation}
where $\alpha_J$ is the pathloss exponent for the \ac{A2G} links, and $\eta_{\textrm{LoS}}$ and $\eta_{\textrm{NLoS}}$ are the attenuation factors for the LoS and the NLoS links, respectively. Also, the \ac{A2G} channel response $h_{m,l}$ and channel gain $g_{m,l}$ are given by $h_{m,l} = (\sqrt{L_{m,l}})^{-1}$ and $g_{m,l} = |h_{m,l}|^2$, respectively.

Let $s_m^c$ be the unit-power symbol sent by UAV $m$ to node $l$ through its sub-channel $c$ with power $p_{m}^{c}$. Then, the received signal $y_l^c$ at node $l$ is given by
\begin{align}\label{eq:sig_nodeL}
    y_l^c&= h_{m,l}\sqrt{p_{m}^{c}}s_m^c +  \sum\limits_{\substack{k\in\mathcal{M}\\
                  k \neq m}} h_{k,l}\sqrt{p_{k}^{c}}s_k^c + w,
\end{align}
where $w$ is the \ac{AWGN} of power $N_0$. Then, the received \ac{SINR} at node $l$ from UAV $m$ through channel $c$ is given by
\begin{equation}\label{eq:IoT:SINR_gam}
    \gamma_{m,l}^c = \frac{a_{l,m,c}\gamma_{m}^{c}g_{m,l}}{\sum\limits_{\substack{k\in\mathcal{M}\\
                  k \neq m}} \gamma_{k}^{c}g_{k,l}  + 1},
\end{equation}
where $\gamma_m^c=\tfrac{p_{m}^{c}}{N_0}$ is the transmit \ac{SINR} at \ac{UAV} $m$ in sub-channel $c$. 
Furthermore, no cooperation is considered among eavesdroppers, i.e. they are non-colluding, thus the eavesdropping risk is dominated by the eavesdropper with the strongest received SINR given by
\begin{align}
    \gamma_{m,e*}^c &=  \frac{\gamma_{m}^{c}g_{m,e*}}{\sum\limits_{\substack{k\in\mathcal{M}\\
                  k \neq m}} \gamma_{k}^{c}g_{k,e*}  + 1},\\
    e* &= \argmax_{e\in\mathcal{E}}\left\{ \frac{\gamma_{m}^{c}g_{m,e}}{\sum\limits_{\substack{k\in\mathcal{M}\\
                  k \neq m}} \gamma_{k}^{c}g_{k,e}  + 1} \right\}.
\end{align}
For ease of notation, $\gamma_{m,l}^c$ will be written as $\gamma_{l}$ when $a_{l,m,c}=1$, and its corresponding $\gamma_{m,e*}^c$ will be written as $\gamma_{e*}$.

The secrecy capacity $C_S$ of the wiretap channel~\cite{Wyner_wiretap}, which is the maximum achievable secrecy rate for a wiretap channel, is defined as $C_S = \left[C_{\mathrm{M}} - C_{\mathrm{W}}\right]^+$~\cite{Cheong_GaussianWiretap} with $[X]^+=\max[X,0]$. Here $C_{\mathrm{M}}$ is the main channel capacity between the legitimate receiver and the legitimate transmitter, and $C_{\mathrm{W}}$ is the wiretap channel capacity between the eavesdropper and the legitimate transmitter. Then, the secrecy capacity for the downlink communication of the corresponding UAV to node $l$, considering Gaussian channels, is given as 
\begin{align}
    \label{eq:secCapacity_b} C_S &=  \left[\log_2\left( \frac{1 + \gamma_{l}}{1 + \gamma_{e*}}  \right) \right]^+.
\end{align}



\section{Sum Secrecy Rate Maximization}

In this section, the optimal node association, the 3D-deployment of UAVs, and the power allocation are obtained to maximize the downlink sum secrecy rate of ground IoT nodes. Considering that the achievable secrecy rate for the node $l$ is given by \eqref{eq:secCapacity_b}, the optimization problem can be formulated as
\begin{subequations}\label{eq:optProbMain_1}
\begin{alignat}{3}\label{eq:optProb_1}
 \mathrm{\textbf{P}:}\;\;\;&\!\max_{\mathbf{A},\{\mathbf{x}_m\}_{m\in \mathcal{M}},\mathbf{P}}        & & \sum\limits_{l\in\mathcal{L}}  \log_2\left( \tfrac{1 + \gamma_{l}}{1 + \gamma_{e*}}\right)  \\
\nonumber &\text{s.t.} &      & \eqref{eq:optProbconst1_1}, \eqref{eq:optProbconst2_1},  \\
\label{eq:optProbconst3_1}&                  &      & a_{l,m,c}  \in \{0,1\}, \quad\quad \forall a_{l,m,c} \in\mathbf{A}\\ 
\label{eq:optProbconst4_1}&                  &      & x_{\mathrm{min}} \leq x_m \leq x_{\mathrm{max}},  \quad\quad \forall m\in\mathcal{M}\\ 
\label{eq:optProbconst5_1}&                  &      & y_{\mathrm{min}} \leq y_m \leq y_{\mathrm{max}},  \quad\quad \forall m\in\mathcal{M}\\ 
\label{eq:optProbconst6_1}&                  &      & z_{\mathrm{min}} \leq z_m \leq z_{\mathrm{max}}, \quad\quad  \forall m\in\mathcal{M}\\
\label{eq:optProbconst7_1}&                  &      &\sum\limits_{c\in\mathcal{C}}p_m^c \leq P, \quad\quad \forall m\in\mathcal{M}.
\end{alignat}
\end{subequations}


While the main goal of solving the optimization problem in~\eqref{eq:optProbMain_1} is to maximize the sum secrecy rate of the system, it is worth noting that not every node will be able to obtain a positive secrecy rate. This occurs because of the high density of eavesdroppers and legitimate nodes present in the system. 
Note that the objective function~\eqref{eq:optProb_1} is a non convex function, and~\eqref{eq:optProbconst3_1} is a mixed-integer constraint, thus problem \textbf{P} is an intricate non-convex combinatorial optimization problem. Alternatively, a block coordinate ascend (BCA) algorithm is proposed to optimize the node association, UAV positioning, and power allocation, each block optimized by considering the other blocks fixed. The proposed secure BCA framework is described next, where each block is optimized at a time while maintaining the others fixed.

\subsection{Node Association}

The first stage consists of solving the optimal association of legitimate IoT nodes to the \ac{UAV}s. Thus, the goal of this stage is to solve the following optimization sub-problem
\begin{alignat}{3}\label{eq:optProb_Ass}
\mathrm{\textbf{P1}:}\;\;\;\;\; &\max_{\mathbf{A}}       &\quad& \sum\limits_{l\in\mathcal{L}} \phi_l &\qquad\qquad \\ 
\nonumber &\text{s.t.} &\qquad&  \eqref{eq:optProbconst1_1}, \ \eqref{eq:optProbconst2_1}, \ \eqref{eq:optProbconst3_1}.
\end{alignat}
Herein, the metric $\phi_l$ is taken, for simplicity, in the high SINR regime by omitting the 1 terms of \eqref{eq:secCapacity_b}, and is given by 
\begin{align}
    \label{eq:phijexp}\phi_l&=\log_2\left(\frac{\frac{g_{m,l}}{I_{m,l}^c  + 1}}{\frac{g_{m,e*}}{ I_{m,e*}^c + 1}}\right)\\
    \label{eq:locInteq}I_{m,n}^c &= \sum\limits_{\substack{k\in\mathcal{M}\\ k \neq m}} \gamma_{k}^{c}g_{k,n},\quad\quad n\in\{l,e*\}.
\end{align}
Note that the power allocated by the UAVs to their subchannels is not considered for the optimization at this stage, thus allowing users to associate based on the channels that offer better secrecy performance.

To solve  \textbf{P1}, a potential game is formulated as described next.

\subsubsection{Potential Game}

According to this game, a fixed number of resources $r_l$, i.e. sub-channels, are available at each \ac{UAV}. By associating to a given resource, a node $l$ will obtain a certain $\phi_l(r_l)$ value, and the goal is to get the highest possible value. 
However, resources are limited, and if a given resource is already occupied, it cannot be assigned to another node. Therefore, there exist a competition among nodes for a given resource in order to obtain the best local secrecy performance. This game consider the following elements:
\begin{itemize}
    \item \textbf{Players:} Are the legitimate nodes $l\in\mathcal{L}$.
    \item \textbf{Actions:} Are the resources to associate with, i.e. the pairs $r_l = (m,c)$, with $m\in\mathcal{M}$ and $c\in\mathcal{C}$.
    \item \textbf{Payoffs:} Are the values $f_l(r_l)=\phi_l(r_l)$ obtained after performing an association.
\end{itemize}

Once the goal is to maximize the sum secrecy of the system, the overall utility can be represented as a function of the actions of every node in the system. 
Then, the utility can be expressed as
\begin{align}\label{eq:utility}
F(\mathrm{\textbf{r}}) = F(r_l,\mathrm{\textbf{r}}_{-l}) =  \sum\limits_{\substack{n\in\mathcal{L}\\ n \neq l}} f_n(r_n) + f_l(r_l).
\end{align}
In~\eqref{eq:utility}, $r_l$ represents the current strategy of node $l$, and $r'_l$ represents a potential new strategy to be adopted, such that the change in payoff for the node $l$ is given by $f_l(r'_l) - f_l(r_l)$. 
By assuming constant power over the association phase, the choice of resource of a given node during this phase does not consider the signal or interference levels at the other nodes, thus $f_n(r_n)$ remains constant under a change of strategy of node $l\neq n$, and then
\begin{align}
\nonumber    &F(r'_l,\mathrm{\textbf{r}}_{-l}) - F(r_l,\mathrm{\textbf{r}}_{-l})    \\
\nonumber&=\left(\sum\limits_{\substack{n\in\mathcal{L}\\ n \neq l}} f_n(r_n) + f_l(r'_l)\right) - \left(\sum\limits_{\substack{n\in\mathcal{L}\\ n \neq l}} f_n(r_n) + f_l(r_l)\right) \\
    &= f_l(r'_l) - f_l(r_l).
\end{align}

This indicates that this is a \textbf{potential game} with the potential function being the overall utility of the system $F(\cdot)$. Therefore, the best response dynamics can be used to reach a pure Nash equilibrium. Furthermore, given that every node can be considered an independent entity, the overall game is a simultaneous move game, where every node chooses its next strategy independently. 

Under these considerations, two conflicts may arise. Particularly, it is possible for more than one node to choose the same resource at a certain moment, and it is also possible for a node to choose an already occupied resource at a certain moment. 
To address these conflicts, it is proposed a protocol to be followed by each UAV. For the first conflict, UAVs will be programmed to allocate the resource to the contending node with the highest $\phi_l$, and if there are two or more nodes with the same value of $\phi_l$, the UAV will associate to one of them arbitrarily. To address the second conflict, nodes are only allowed to choose resources that are not currently occupied. It can be seen as the UAVs advertising only their available sub-channels to the legitimate nodes.

Apart from best response dynamics, a potential game is guaranteed to reach a pure Nash equilibrium under a \ac{SLLL} algorithm~\cite{MardenLLL}, which is described next.

\subsubsection{Synchronous Log Linear Learning}

In this algorithm, it is considered that the gain in payoff, obtained by performing an action, changes with respect to the current action (marginal payoff), which is given by
\begin{equation}\label{eq:marginalPayoff_SLLL}
    f_l(r'_l)= \phi_l(r'_l)-\phi_l(r_l).
\end{equation}
Therefore, the gain in payoff obtained by remaining in the current strategy is 0 and the potential game modeling holds.

The \ac{SLLL} algorithm 
is considered for the potential game with 
\eqref{eq:marginalPayoff_SLLL} as the payoff function. Under the \ac{SLLL} algorithm, a legitimate node chooses an action from their available actions following the \ac{SBR} mixed strategy~\cite{HasanbeigSBR} given by
\begin{equation}\label{eq:softmax}
    \pi_l(r_l) = \frac{e^{f_l(r_l)}}{\sum\limits_{z_l\in\mathcal{A}_l}e^{f_l(z_l)}}.
\end{equation}
After each legitimate node has chosen an action, if two or more nodes choose the same resource, UAVs apply the protocol to solve conflicts, then all the legitimate nodes choose their next strategy. This goes on until no legitimate nodes have available strategies, i.e., until no node has an incentive to change strategies (i.e., they are already in their best response strategy), which constitutes a pure Nash equilibrium. Algorithm \ref{alg:SLLL_Ass} describes the operation of this algorithm.

\begin{algorithm}[h!]\footnotesize
\caption{SLLL for node association algorithm}\label{alg:SLLL_Ass}
counter $\; \leftarrow \; 0$\;
\While{counter $<$ n\_iter }{
    conv\_flag $\; \leftarrow \; 1$ \;
    $\mathrm{\textbf{x}}[l]\; \leftarrow \; -1\;\;\forall l\in\mathcal{L}$\;
    \For{$l\in \mathcal{L}$}{
        $\mathcal{A}_l \; \leftarrow \; \{r=(m,c), \;\;s.t.\;\; (m,c)\in\mathcal{M}\times\mathcal{C}$\}\;
        $\mathcal{A}_l \; \leftarrow \; \mathcal{A}_l\setminus\{ r=(m,c), \;\;s.t.\;\;  \sum_{n\in \mathcal{L}}a_{n,m,c}>0 \}$\;
        
        
        $f_l(r)\; \leftarrow \; $ compute as in \eqref{eq:marginalPayoff_SLLL} $\forall r \in \mathcal{A}_l$\;
        $\mathcal{A}_l \; \leftarrow \; \mathcal{A}_l\setminus\{ r=(m,c) \;\;s.t.\;\; f_l(r)\leq0\}$ \;
        \uIf{ $\mathcal{A}_n \neq \emptyset$}{
            $\mathrm{Pr}\left[ X_l = r \right]\; \leftarrow \; $ compute as in \eqref{eq:softmax} $\forall r \in \mathcal{A}_l$\; \label{alg:in:softmaxEnf}
            $x_l \; \leftarrow \; $ choose from $r\in\mathcal{A}_l$ according to $\mathrm{Pr}\left[ X_l = r \right]$\; 
            $\mathrm{\textbf{x}}[l]\; \leftarrow \;x_l$\;
            conv\_flag $\; \leftarrow \; 0$ \;
        }
    }
    \uIf{ conv\_flag $\; == \; 1$ }{
        Stop the association process\;
    }
    
    \For{$m\in \mathcal{M}$}{
        \For{$c\in \{c\in\mathcal{C}\;\; s.t. \sum_{n\in \mathcal{L}}a_{n,m,c}=0 \} $}{
            $\mathcal{N}_{m,c} \; \leftarrow \; \{ l \;\;s.t.\;\; \mathrm{\textbf{x}}[l] =   (m,c)$\}\;
            \uIf{ $|\mathcal{N}_{m,c}| > 0$}{
                $f_l(m,c)\; \leftarrow \; $ compute as in \eqref{eq:marginalPayoff_SLLL} $\forall l\in \mathcal{N}_{m,c}$\;
                $f_{l,\mathrm{max}}\; \leftarrow \; \max_{l\in \mathcal{N}_{m,c}} f_l(m,c)$ \;
                $\mathcal{N}_{m,c,\mathrm{max}}\; \leftarrow \;\{ l\in \mathcal{N}_{m,c} \;\;s.t.\;\; f_l(m,c) = f_{l,\mathrm{max}} \}$\;
                $l^*\; \leftarrow \;$ choose from $l\in \mathcal{N}_{m,c,\mathrm{max}}$ randomly\;
                $(m_{\mathrm{prev}},c_{\mathrm{prev}})  \; \leftarrow \;   (m,c) \;\;s.t.\;\; a_{l^*,m,c}=1$\;
                $a_{l^*,m_{\mathrm{prev}},c_{\mathrm{prev}}}  \; \leftarrow \;   0$\;
                $a_{l^*,m,c}  \; \leftarrow \;   1$\;
            }
        }
    }
    counter $\; \leftarrow \; $counter + 1\;
    
}
\end{algorithm}

\subsection{UAV Position Control}

The second stage in the framework consists of the 3D positioning of the \ac{UAV}s within region $S$ based on the sum secrecy rate obtained by each UAV, having $\mathcal{L}_m$ be the set of legitimate nodes associated to UAV $m$. 

For the UAV positioning, the following optimization subproblem is formulated
\begin{subequations}\label{eq:optProbMain_Pos}
\begin{alignat}{3}\label{eq:optProb_Pos}
\mathrm{\textbf{P2}:}\;\;\;\;\;\;\;\;\; &\!\max_{\mathbf{A},\{\mathbf{x}_m\}_{m\in \mathcal{M}}}        &\qquad& \Phi = \sum\limits_{l\in\mathcal{L}} \phi_l &\qquad\qquad \\ \label{eq:optProbconst1_Pos}
&\text{s.t.} &      & \eqref{eq:optProbconst4_1},\eqref{eq:optProbconst5_1},\eqref{eq:optProbconst6_1}.   &
\end{alignat}
\end{subequations}

The positioning of the \ac{UAV}s, unlike the association of the nodes, is performed over a continuous domain which is the entire region, with a continuous altitude range, for all of the \ac{UAV}s. Heuristic methods have shown to work well over a continuous space,  such as particle swarm optimization~\cite{PSO_REF} and genetic algorithm~\cite{GENALGO_REF}. However, these methods require increased complexity, continuous coordination between the agents, and longer convergence time. While the outcomes from these continuous-domain algorithms are close to optimum values, discrete-domain algorithms may provide simpler and satisfactory solutions, which is beneficial when considering resource-limited IoT nodes.

Thus, a two-stage positioning protocol is proposed, where a global 2D $M$-centroid clustering is solved as the first stage, then an individual altitude selection is performed over the altitude range $\Delta z$ discretized over $N_z$ altitude levels. The set of discretized altitude levels is denoted as $\mathcal{Z}$, with $|\mathcal{Z}| = N_z$.  The two stages of this protocol are described in the following.

\subsubsection{2D Clustering}

\begin{figure}[ht]
\centering
\resizebox{\columnwidth}{!}{
\begin{tikzpicture}[node distance = 2cm, scale=0.85, every node/.style={scale=0.85}]

\node (start)       [startstop] {\textbf{Start}};
\node (inAll)       [process, below of= start, text width = 5cm]      {$\forall c\in\{1,...,K\}$\\Initiate $(x_c,y_c)$ randomly  };
\node (node1)     [process, below of= inAll, text width = 5cm]      {$i=1$\\$\forall c\in\{1,...,K\} \;\;C_c = \emptyset $};
\node (node2a)     [process, below of= node1, text width = 7cm]      {$c^*=\argmin_{c}\sqrt{(x_i-x_c)^2 + (y_i-y_c)^2}$\\ $C_{c^*} = C_{c^*} \cup \{i\} $};
\node (node2)     [process, below of= node2a]      {$i=i+1$};
\node (node3)     [decision, right of= node2, xshift=3cm]      {$i\leq L$};
\node (moveCent)    [process, below of= node2]    {$\forall c\in\{1,...,K\}$\\ $x_c = \frac{1}{|C_c|}\sum_{i\in C_c}x_i$\\$y_c = \frac{1}{|C_c|}\sum_{i\in C_c}y_i$\\};
\node (conv)        [decision, left of= moveCent, xshift=-3.5cm]   {Centroids move?};
\node (end)         [startstop, below of= conv, yshift=-1cm]     {\textbf{Stop}};

\draw[arrow] (start) -- (inAll);
\draw[arrow] (inAll) -- (node1);
\draw[arrow] (node1) -- (node2a);
\draw[arrow] (node2a) -- (node2);
\draw[arrow] (node2) -- (node3);
\draw[arrow] (node3) --node[anchor=west] {Yes}  ++(0,2) |- (node2a);
\draw[arrow] (node3) --node[anchor=west] {No}  ++(0,-2) |- (moveCent);
\draw[arrow] (moveCent) -- (conv);
\draw[arrow] (conv) --node[anchor=east] {No}  (end);
\draw[arrow] (conv) --node[anchor=east] {Yes}  ++(0,2) |- (node1);

\end{tikzpicture}
}
\caption{K-means algorithm flowchart}
\label{flow:kmeans}
\end{figure}
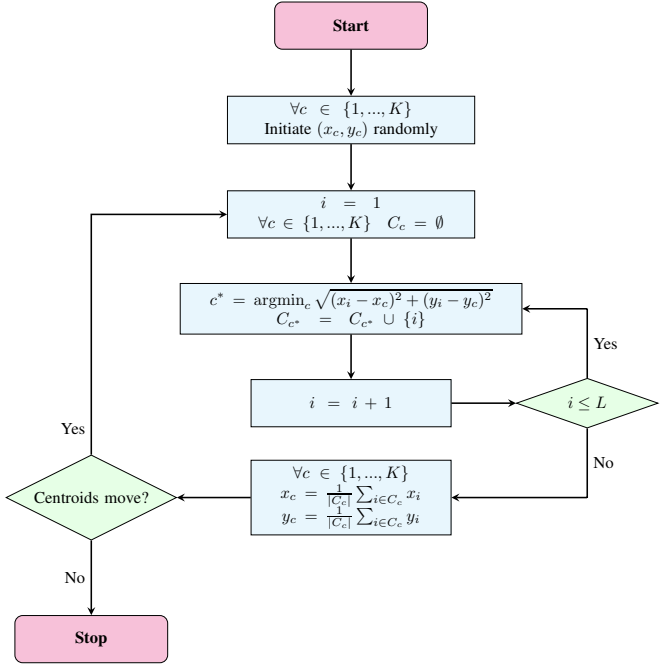

For the 2D positioning, we aim at finding the 2D points with the highest concentration of legitimate nodes, or barycenters of the concentrations of nodes, which will privilege the best secrecy coverage. For this purpose, the unsupervised learning algorithm k-means clustering~\cite{kmeans_REF} is applied, which returns the centroids of the clusters (points in the area) and the members of each cluster. A diagram of this algorithm can be seen in Fig.~\ref{flow:kmeans}.

The k-means algorithm requires the knowledge of the position of the legitimate nodes of the system. Then, the algorithm is run at some central unit (one of the UAVs) only once for the real positions of the nodes.


\subsubsection{Best Response Dynamics}

Once the UAV 2D positioning is solved, the UAV altitude selection problem can be formulated as a game consisting of
\begin{itemize}
    \item \textbf{Players: } UAVs $m\in\mathcal{M}$.
    \item \textbf{Actions: } discrete altitude levels $r_m = z_m\in\mathcal{Z}$.
    \item \textbf{Payoffs: } the sum secrecy metric obtained by their associated nodes $f_m(r_m)=\Phi_m(r_m)=\sum_{l\in\mathcal{L}_m}\phi_l$.
\end{itemize}

We utilize a best response algorithm to solve the positioning problem with a modified payoff into the marginal gain payoff of UAV $m$ for choosing altitude $r_m'$:
\begin{equation}\label{eq:payoff_BR_Pos}
    f_m(r_m')=\Phi_m(r_m') - \Phi_m(r_m).
\end{equation}
where $r_m$ is the current position of UAV $m$. Then this algorithm considers the simple action selection per UAV, i.e. $r_m = \argmax_{z_m\in\mathcal{Z}} z_m$, 
which is performed simultaneously and independently at each UAV. This algorithm is described at Algorithm \ref{alg:BR_Pos}.

\begin{algorithm}[h!]\footnotesize
\caption{Best response for UAV altitude positioning algorithm}\label{alg:BR_Pos}
counter $\; \leftarrow \; 0$\;
\While{counter $<$ n\_iter }{
    conv\_flag $\; \leftarrow \; 1$ \;
    \For{$m\in \mathcal{M}$}{
        $\mathcal{A}_m \; \leftarrow \; \mathcal{Z}$\;
        $f_m(r)\; \leftarrow \; $ compute as in \eqref{eq:payoff_BR_Pos} $\forall r \in \mathcal{A}_m$\;
        $\mathcal{A}_m \; \leftarrow \; \mathcal{A}_m\setminus\{ r\in\mathcal{A}_m \;\;s.t.\;\; f_m(r)\leq0\}$ \;
        \uIf{ $\mathcal{A}_m \neq \emptyset$}{
            $f_{m,\mathrm{max}}\; \leftarrow \; \max_{r\in \mathcal{A}_m} f_m(r)$ \;
            $\mathcal{A}_{m,\mathrm{max}}\; \leftarrow \;\{ r\in \mathcal{A}_m \;\;s.t.\;\; f_m(r) = f_{m,\mathrm{max}} \}$\;
            $z_m\; \leftarrow \;$choose from $r\in \mathcal{A}_{m,\mathrm{max}}$ randomly\;
            conv\_flag $\; \leftarrow \; 0$ \;
            Make UAV $m$ assume altitude $z_m$\;
        }
    }
    \uIf{ conv\_flag $\; == \; 1$ }{
        Stop the positioning process \;
    }
    counter $\; \leftarrow \;$ counter + 1\;
}
\end{algorithm}


The information required for Algorithm \ref{alg:BR_Pos} is local to each UAV, disregarding the strategy taken by other UAVs or their exact positions. This algorithm is fast compared to exhaustive search, and it usually converges within two or three iterations.

\subsection{Secure Power Allocation}

In the third and final stage, each UAV allocates its available power to the nodes associated to them. To this end, the following convex optimization problem is addressed
\begin{subequations}\label{eq:optProb_Pow2_Init}
\begin{alignat}{3}\label{eq:optProb_Pow}
\mathrm{\textbf{P3}:}\;\;\;\;\; &\max_{\mathbf{P}}       &\quad& \sum\limits_{l\in\mathcal{L}} \log_2\left( \tfrac{1 + \gamma_{l}}{1 + \gamma_{e*}}\right) &\qquad \\ 
\nonumber &\text{s.t.} &\qquad&  \eqref{eq:optProbconst7_1}.
\end{alignat}
\end{subequations}

In \textbf{P3}, the objective \eqref{eq:optProb_Pow} is non-convex on $\mathbf{P}$, so this problem cannot be directly solved. Moreover, the condition for secrecy for a user is given by
\begin{equation}\label{eq:guaranteeSec}
    \frac{g_{m,l}}{I_{m,l}^c  + 1} > \frac{g_{m,e*}}{ I_{m,e*}^c + 1},
\end{equation}
which cannot be guaranteed to all nodes. In that case, the power optimization formulation as expressed in~\textbf{P3} will allocate all the power budget only to the nodes that can achieve secrecy, leaving without power to those that cannot, which is not desirable. Alternatively, it is considered to the original problem in order to guarantee a minimum SINR requirement to every node in the system. To that purpose, the set $\mathcal{L}_m^{S}$ is introduced as the set of nodes associated to UAV $m$ that can be guaranteed secrecy, that is to say, for which \eqref{eq:guaranteeSec} holds. Afterwards, the proposed optimization problem is a max-min secrecy rate problem for the nodes in $\mathcal{L}_m^{S}$, performed locally at each UAV, expressed as
\begin{subequations}\label{eq:optProb_Pow2_Main_pre}
\begin{alignat}{3}\label{eq:optProb_Pow2_pre}
\;\;\; &\max_{\mathbf{p}_m} \min_{l\in\mathcal{L}_m^{S}} & &\log_2\left( \tfrac{1 + \gamma_{l}}{1 + \gamma_{e*}}\right) & \\ 
\label{eq:optProbconst1_Pow2_pre} &\text{s.t.} &\quad&  \gamma_{l} > \gamma_0 & \forall l\in\mathcal{L}_m \\
\label{eq:optProbconst3_Pow2_pre} & &\quad&  \sum\limits_{c\in\mathcal{C}}p_m^c \leq P, & 
\end{alignat}
\end{subequations}

An equivalent optimization problem can be formulated as
\begin{subequations}\label{eq:optProb_Pow2_Main}
\begin{alignat}{3}\label{eq:optProb_Pow2}
\mathrm{\textbf{P3'}:}\;\;\; &\max_{\mathbf{p}_m} & &R_S & \\ 
\label{eq:optProbconst1_Pow2} &\text{s.t.} &\quad&  \gamma_{l} > \gamma_0 & \forall l\in\mathcal{L}_m \\
\label{eq:optProbconst2_Pow2} & &\quad&  \log_2\left( \tfrac{1 + \gamma_{l}}{1 + \gamma_{e*}}\right) > R_S &\quad \forall l\in\mathcal{L}_m^{S} \\
\label{eq:optProbconst3_Pow2} & &\quad&  \sum\limits_{c\in\mathcal{C}}p_m^c \leq P, & 
\end{alignat}
\end{subequations}
In this formulation, the interference perceived at each node is assumed constant over the optimization process, and an iterative optimization scheme can be applied. Thus, the interference at its associated nodes are computed at each UAV, and problem \textbf{P3'} is solved in parallel in all UAVs. Then the updated interference terms are computed, and the process is repeated until convergence or for a number of iterations. 

Once \textbf{P3'} is convex, it can be split into two subproblems, \textbf{P3'a} and \textbf{P3'b}, as 
\begin{subequations}\label{eq:optProb_Pow2a_Main}
\begin{alignat}{3}\label{eq:optProb_Pow2a}
\mathrm{\textbf{P3'a}:}\;\;\; &\min_{\mathbf{p}_m^{(a)}} & &P_{NS} & \\ 
\label{eq:optProbconst1_Pow2a} &\text{s.t.} &\quad&  \gamma_{l} > \gamma_0 &\quad \forall l\in\mathcal{L}_m. \qquad \qquad \quad
\end{alignat}
\end{subequations}

\begin{subequations}\label{eq:optProb_Pow2b_Main}
\begin{alignat}{3}\label{eq:optProb_Pow2b}
\mathrm{\textbf{P3'b}:}\;\;\; &\max_{\mathbf{p}_m^{(b)}} & &R_S  \\ 
\label{eq:optProbconst1_Pow2b} &\text{s.t.} & &  \log_2\left( \tfrac{1 + \gamma_{l}}{1 + \gamma_{e*}}\right) > R_S \quad \forall l\in\mathcal{L}_m^{S} \\
\label{eq:optProbconst2_Pow2b} & &\quad&  \sum\limits_{c\in\mathcal{C}}p_m^{c,(b)} \leq P_{S},
\end{alignat}
\end{subequations}
where $\mathbf{p}_m^{(a)}$ is the power profile for the minimum SINR requirement, and $\mathbf{p}_m^{(b)}$ is the power profile for the max-min secrecy rate optimization, such that $\mathbf{p}_m = \mathbf{p}_m^{(a)}+\mathbf{p}_m^{(b)}$, $P_{NS}$ is the power used to meet the minimum SINR requirement, and $P_{S}=\left[P-P_{NS}\right]^+$ is the power available for max-min secrecy rate optimization.  

First, problem \textbf{P3'a} is solved for the power profile $\mathbf{p}_m^{(a)}$ and power $P_{NS}$ is found, which is power required to guarantee the minimum SINR $\gamma_0$ for all associated nodes. If $P_{NS}\geq P$, there is not enough power to meet the SINR constraint, then the overall local power profile is taken as $\mathbf{p}_m = \mathbf{p}_m^{(a)}(P/P_{NS})$, and the local power allocation process ends. If $P_{NS} < P$, then the available power for the max-min secrecy rate problem is assumed as $P_{S} = P - P_{NS}$, and the problem \textbf{P3'b} is solved by obtaining the power profile $\mathbf{p}_m^{(b)}$, and the overall local power profile is given as $\mathbf{p}_m = \mathbf{p}_m^{(a)}+\mathbf{p}_m^{(b)}$.

The closed form solution for problem \textbf{P3'a} is given as
 \begin{equation}\label{eq:powAexp}
     p_m^{c,(a)} = \gamma_0 \left( \frac{I_{m,l}^c  + 1}{g_{m,l}}\right) \quad \forall l\in \mathcal{L}_m
 \end{equation}

Problem \textbf{P3'b} can be solved by bisection over the following minimum power optimization problem
\begin{subequations}\label{eq:optProb_Pow2bMinPow_Main}
\begin{alignat}{3}\label{eq:optProb_Pow2amp}
\mathrm{\textbf{P3'b'}:}\;\;\; &\min_{\mathbf{p}_m^{(b)}} & &P_{S} & \\ 
\label{eq:optProbconst1_Pow2amp} &\text{s.t.} &\quad&   \tfrac{1 + \gamma_{l}}{1 + \gamma_{e*}} > \gamma_S &\quad \forall l\in\mathcal{L}_m^S.
\end{alignat}
\end{subequations}
where $\gamma_S=2^{R_S}$. This problem has the following closed-form solution
\begin{equation}\label{eq:closedFormgamS}
    p_m^{c,(b)} = \left[ \frac{\gamma_S - 1}{\frac{g_{m,l}}{I_{m,l}^c  + 1} - \gamma_S\left( \frac{g_{m,e*}}{ I_{m,e*}^c + 1} \right)} \right]^+\quad \forall m\in \mathcal{L}_m.
\end{equation}
Considering that this problem is solved for nodes that can achieve secrecy, and assuring that $p_m^{c,(b)}$ is non-zero, the bounds for $\gamma_S$ are
\begin{equation}\label{eq:gamSbounds}
    1 < \gamma_S < \min_{l\in\mathcal{L}_m^S}\left\{ \frac{\frac{g_{m,l}}{I_{m,l}^c  + 1}}{\frac{g_{m,e*}}{ I_{m,e*}^c + 1}} \right\}
\end{equation}

All in all, to solve problem \textbf{P3'b}, bisection is performed on problem \textbf{P3'b'} with closed form solution \eqref{eq:closedFormgamS}, over $\gamma_S$, whose initial minimum and maximum values are given by the bounds in \eqref{eq:gamSbounds}. The power allocation algorithm is described in Algorithm \ref{alg:Pow}.

\begin{algorithm}[h!]\footnotesize
\caption{Secure power allocation algorithm}\label{alg:Pow}

\While{counter $<$ n\_iter\_pow}{
    \For{$m\in\mathcal{M}$}{
        \For{$l\in\mathcal{L}_m$}{
            $I_{m,l} \; \leftarrow \; $  compute as in \eqref{eq:locInteq}\;
            $I_{m,e*} \; \leftarrow \; $ compute as in \eqref{eq:locInteq}\;
        }
    }
    \For{$m\in\mathcal{M}$}{
        $\mathcal{L}_m^S \; \leftarrow \; \{\}$\;
        \For{$l\in\mathcal{L}_m$}{
            \uIf{ \eqref{eq:guaranteeSec} holds}{
                $\mathcal{L}_m^S \; \leftarrow \; \mathcal{L}_m^S \cup \{l\}$\;
            }
        }
        \For{$l\in\mathcal{L}_m$}{
            $p_m^{c,(a)} \; \leftarrow \; $ compute as in \eqref{eq:powAexp}\;
        }
        $P_{NS} \; \leftarrow \; \sum_{l\in\mathcal{L}_m} p_m^{c,(a)}$\;
        \uIf{$P_{NS} \geq P$ \textbf{OR} $\mathcal{L}_m^S$ is empty}{
            \For{$l\in\mathcal{L}_m$}{
                $p_m^{c} \; \leftarrow \; p_m^{c,(a)}(P/P_{NS})$ \;
            }
            \textbf{continue}\;
        }
        $P_{S} \; \leftarrow \; P-P_{NS}$\;
        $\gamma_{\mathrm{min}}, \gamma_{\mathrm{max}} \; \leftarrow \; $ set according to \eqref{eq:gamSbounds}\;
        \While{counter\_bis $<$ n\_iter\_bis}{
            $\gamma_S \; \leftarrow \; \frac{1}{2}(\gamma_{\mathrm{min}} + \gamma_{\mathrm{max}} ) $\;
            \For{$l\in\mathcal{L}_m^S$}{
                $p_m^{c,(a)} \; \leftarrow \; $ compute as in \eqref{eq:closedFormgamS}\;
            }
            \uIf{$\sum_{l\in\mathcal{L}_m} p_m^{c,(b)} > P_S$}{
                $\gamma_{\mathrm{max}} \; \leftarrow \; \gamma_S$\;
            }
            \uIf{$\sum_{l\in\mathcal{L}_m} p_m^{c,(b)} < P_S$}{
                $\gamma_{\mathrm{min}} \; \leftarrow \; \gamma_S$\;
            }
        }
        \For{$l\in\mathcal{L}_m^S$}{
            $p_m^{c} \; \leftarrow \; p_m^{c,(a)} + p_m^{c,(b)}$\;
        }
    }
    counter $\; \leftarrow \; $counter + 1\;
}

\end{algorithm}

\section{Results and Discussion}

In this section, the performance of the proposed framework is evaluated through Monte Carlo simulations. For that purpose, unless otherwise stated, the adopted simulation parameters are presented in Table~\ref{tab:simParam}. Therein, $\gamma_P=P/N_0$ is the total transmit SNR of each UAV, and $N_{it}$ is the number of association-positioning iterations for a given realization of the system. 
The number of UAVs $M$ to be deployed is chosen such that $(M-1)C < L\leq MC$.

\begin{table}[bt]
    \centering
    \begin{tabular}{|c|c||c|c|}\hline
         \textbf{Parameter} & \textbf{Value}& \textbf{Parameter} & \textbf{Value}  \\\hline
        $N$                    &80           & $C$                            &8\\ \hline
         $N_{it}$                 &5          & $\gamma_P$                        &20 $dB$\\ \hline
         $x_{\mathrm{min}}$     &0 m        & $q$                            &0.5\\ \hline
         $x_{\mathrm{max}}$     &1000 m     & $\psi$ (Urban)                 & 9.61 \\ \hline
         $y_{\mathrm{min}}$     &0 m        & $\omega$ (Urban)                & 0.16\\ \hline
         $y_{\mathrm{max}}$     &1000 m     & $\eta_{\mathrm{LoS}}$ (Urban)   & 1.0\\ \hline
         $z_{\mathrm{min}}$     &20 m       & $\eta_{\mathrm{NLoS}}$ (Urban)  & 20\\ \hline
         $z_{\mathrm{max}}$     &300 m      & $\alpha_G$ (Urban)              & 0.3\\ \hline
         $N_z$                  &8          & $\alpha_J$ (Urban)              & 0.3\\ \hline
    \end{tabular}
    \caption{Monte Carlo simulations common parameters.}
    \label{tab:simParam}
\end{table}

Unless otherwise stated, for each realization the following steps are taken
\begin{enumerate}
    \item The $N$ nodes are distributed over the region following a binomial point process.
    \item Legitimate nodes are selected following a Bernoulli distribution of parameter $q$.
    \item The association and positioning processes are performed subsequently a number $N_{it}$ of iterations.
\end{enumerate}

\subsection{Association and Positioning Benchmarks}

Three association and positioning benchmarks are presented for the sake of comparison: {\color{black}
\begin{enumerate}
    \item \textbf{Best Response Association: } Framework with a best response algorithm for the association phase. Similar to Algorithm \ref{alg:SLLL_Ass}, but on line \ref{alg:in:softmaxEnf}, $\mathrm{Pr}\left[ X_l = r \right]=1$ for $r = \argmax_{r_l\in\mathcal{A}_l} f_l(r_l)$ and zero for the rest of available actions $r_l\in\mathcal{A}_l\setminus\{r\}$.
    \item \textbf{Greedy Association: } Framework with greedy association algorithm from \cite{HammoutiGreedyAlgsUAV}. This approach iteratively associates the best node-UAV pair through the system in terms of the secrecy rate, until all nodes are associated. 
    \item \textbf{Adapted Greedy: }  Framework with adapted greedy algorithm for association and positioning from \cite{HammoutiGreedyAlgsUAV}. This approach positions each UAV one by one, and associates to it the nodes that present the best secrecy rate, until all UAVs are positioned, and all nodes associated.
\end{enumerate} }


\begin{figure}[h!]
    \centering
    \includegraphics[width=0.75\linewidth]{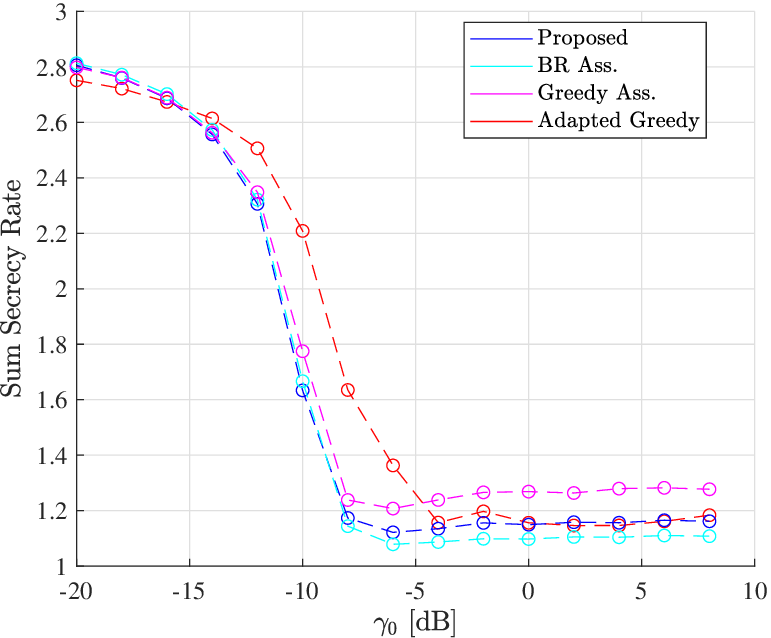}
    \caption{Average sum secrecy rate vs. minimum SINR constraint $\gamma_0$ obtained by different frameworks.}
    \label{fig:gam0_sumcs}
\end{figure}

Fig.~\ref{fig:gam0_sumcs} shows the sum secrecy rate of the system versus $\gamma_0$ for the proposed secure power allocation scheme, and results are compared to the benchmarks described above. It can be seen that, for smaller $\gamma_0$ values, where more power is allocated for the max-min secrecy rate subproblem, the proposed framework and the one with best response association perform better than the greedy benchmarks. On the other hand, for larger $\gamma_0$ values, where the power allocation tends to a max-min SINR power allocation, the proposed solution performs better than the one with best response association, as good as the one for adapted greedy benchmark, but worse than the greedy association benchmark.

\begin{figure}[h!]
    \centering
    \includegraphics[width=0.75\linewidth]{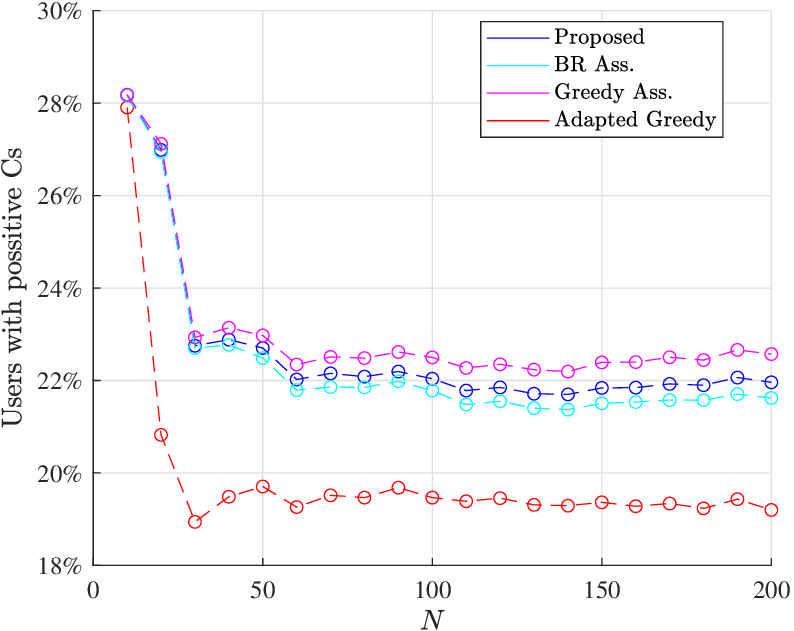}
    \caption{Average percentage of legitimate nodes with positive secrecy rate vs. number of IoT nodes in the system $N$, obtained by different frameworks.}
    \label{fig:N_numcs}
\end{figure}

Fig.~\ref{fig:N_numcs} shows the percentage of legitimate nodes that achieve positive secrecy rate versus the number of nodes in the system $N$, with $\gamma_0 = -10$dB and $M$ chosen such that $(M-1)C < L\leq MC$. Note that there is an initial drop in the percentage of users with positive secrecy for small $N$ values due to the added interference of an increasing number of UAVs. However, after a certain value of $N$, the percentage of users with positive secrecy in the system remains steady, where the proposed framework performs better than the adapted greedy and best response association benchmarks, but worse than the greedy association benchmark. While the greedy association benchmark outperforms the proposed framework, the greedy association is more complex and presents slow convergence.

The best response association benchmark exhibits a similar complexity than the proposed association solution, the greedy association, the adapted greedy association, while positioning benchmarks have an increased complexity in their executions, require more coordination, and take a longer time to converge. Then, letting $T_{\mathrm{ass}}$ and $T_{\mathrm{pos}}$ be the times for a round of association iterations and of positioning iterations, respectively, and $T_{\mathrm{pow}}$ be the total time of the power allocation. It can be observed that, with $N=80$ for the proposed framework and the framework with best response association, the node association finds a Nash Equilibrium in less than 10 iterations, the UAV positioning finds a Nash Equilibrium in less than 3 iterations, and the overall framework converges in less than 5 iterations. The overall convergence time of the frameworks are presented in Table~\ref{tab:FWtimes}, for $N=80$.

\begin{table}[h!]
    \centering
    \begin{tabular}{|c|c|}\hline
         \textbf{Framework} & \textbf{Convergence Time} \\\hline
        Proposed    & $(10T_{\mathrm{ass}} + 2 T_{\mathrm{pos}} + T_{\mathrm{pow}})5$ \\ \hline
        BR Ass.    & $(10T_{\mathrm{ass}} + 2 T_{\mathrm{pos}} + T_{\mathrm{pow}})5$ \\ \hline
        Greedy Ass.   & $(NT_{\mathrm{ass}} + 2 T_{\mathrm{pos}} + T_{\mathrm{pow}})5$ \\ \hline
        Adapted greedy   & $(N_zT_{\mathrm{ass}} + N_z T_{\mathrm{pos}} ) M + + T_{\mathrm{pow}}$ \\ \hline
    \end{tabular}
    \caption{Convergence times.}
    \label{tab:FWtimes}
\end{table}

Therefore, the proposed framework presents much smaller convergence times than the greedy algorithms presented in \cite{HammoutiGreedyAlgsUAV}, while approaching the greedy association benchmark results.

\subsection{Power Allocation Benchmarks}

To compare the proposed secure power allocation strategy, the following power allocation benchmarks are considered {\color{black}
\begin{enumerate}
    \item \textbf{Max. Min SINR: } An iterative local max-min SINR power allocation per UAV. It solves the following optimization problem
\begin{subequations}\label{eq:BM01_00}
\begin{alignat}{3}\label{eq:BM01_01}
\;\;\; &\max_{\mathbf{p}_m} \min_{l\in\mathcal{L}_m} & &\gamma_{l} & \\ 
\label{eq:BM01_02} &\text{s.t.} &\quad&  \gamma_{l} > \gamma_0 & \forall l\in\mathcal{L}_m \\
\label{eq:BM01_03} & &\quad&  \sum\limits_{c\in\mathcal{C}}p_m^c \leq P, & 
\end{alignat}
\end{subequations}
    This power allocation scheme targets to guarantee the same SINR to all the nodes served by a given UAV.
    \item \textbf{Max. Sum Rate: } An iterative local sum-rate maximization power allocation per UAV. It solves the following optimization problem
\begin{subequations}\label{eq:BM02_00}
\begin{alignat}{3}\label{eq:BM02_01}
\;\;\; &\max_{\mathbf{p}_m}  & & \sum_{ l\in\mathcal{L}_m } \log_2\left(1 + \gamma_{l}\right) & \\ 
\label{eq:BM02_02} &\text{s.t.}  &\quad&  \sum\limits_{c\in\mathcal{C}}p_m^c \leq P, & 
\end{alignat}
\end{subequations}
    This power allocation scheme seeks to maximize the sum rate across all of the nodes served by a UAV. By doing so, it may cause some nodes to have no power allocated to them.
\end{enumerate}}
    The proposed power allocation strategy as well as the power allocation benchmarks are performed with the secure association and positioning phases proposed.

\begin{figure}[h!]
    \centering
    \includegraphics[width=0.75\linewidth]{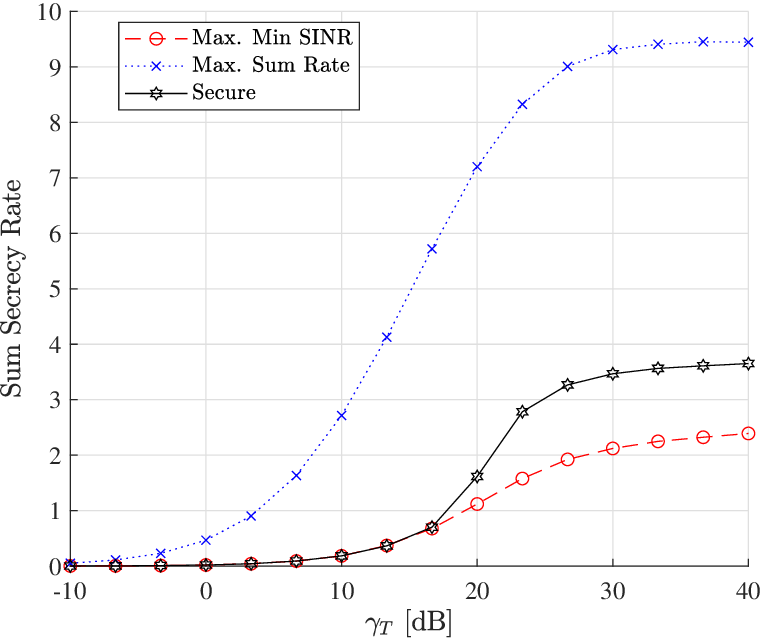}
    \caption{Average sum secrecy rate vs. transmit SNR available to UAVs, obtained by the different power allocation schemes.}
    \label{fig:gamt_sumcs}
\end{figure}

Fig.~\ref{fig:gamt_sumcs} shows the sum secrecy rate of the system versus $\gamma_T$ for the proposed secure power allocation scheme compared to the benchmarks with $\gamma_0 = -10$dB. It can be seen that for smaller transmit SNR values, the proposed power allocation scheme matches with the max-min benchmark.  This behavior occurs because, at these ranges of $\gamma_T$, there is not enough transmit SNR to satisfy the minimum SINR requirement, so no power is allocated for $P_S$. 
At higher $\gamma_T$ values, the proposed scheme outperforms the max-min benchmark, as power is allocated for secrecy improvement after fulfilling the minimum SINR requirements for all nodes. On the other hand, the max. sum rate benchmark outperforms the proposed secure power allocation scheme in terms of sum secrecy rate. However, the max. sum rate scheme allocates all the power of a given UAV to the nodes with the strongest channel to it. This causes the nodes with weaker channels to their serving UAV to receive no power from it, effectively disconnecting a large number of nodes from the network, as can be seen in the next figure.


\begin{figure}[h!]
    \centering
    \includegraphics[width=0.75\linewidth]{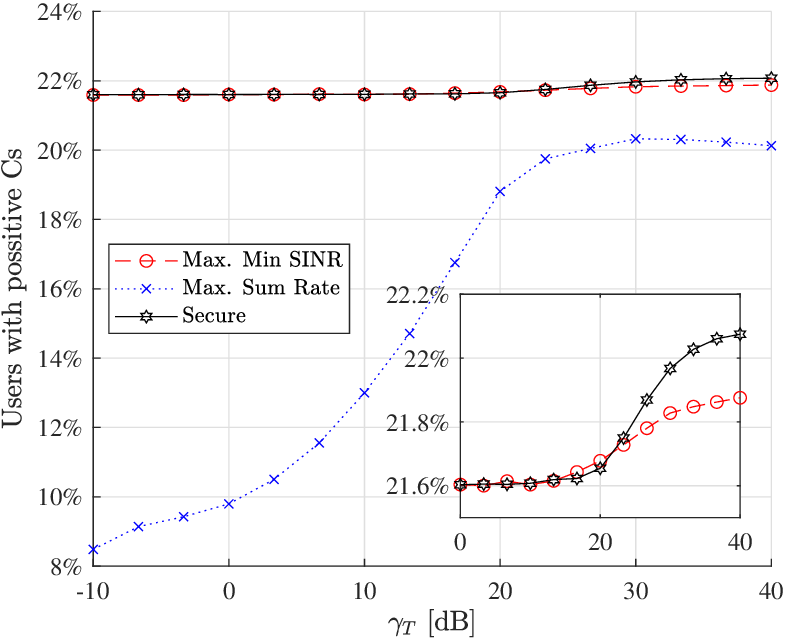}
    \caption{Average percentage of legitimate nodes with positive secrecy rate vs.transmit SNR available to UAVs, obtained by the different power allocation schemes.}
    \label{fig:gamt_numcs}
\end{figure}

Fig.~\ref{fig:gamt_numcs} shows the percentage of legitimate nodes that are able to achieve positive secrecy rate versus $\gamma_T$, for the proposed secure power allocation scheme compared to the benchmarks and $\gamma_0 = -10$dB. Note that the proposed secure power allocation scheme presents a similar behavior compared to the max-min SINR benchmark as in the previous figure. However, it can be seen that the max-sum-rate benchmark presents a significant smaller number of users that can achieve secrecy in the system due to all the power being allocated only to the users with strongest channels. Even for high $\gamma_T$ values, the performance of max-sum-rate benchmark is still worse than the proposed secure power allocation in terms of users that achieve positive secrecy rates in the system.

\section{Conclusions}

In this work, an IoT scenario was investigated, where a swarm of UAVs, acting as ABSs, provide coverage to a group of ground nodes, while considering all nodes that do not participate of the communication process as eavesdroppers. In this scenario, the maximization of the sum-secrecy rate of the system is addressed by proposing a BCA secure framework consisting of the association of the ground nodes, the 3D positioning of the UAVs, and the power allocation for the associated nodes. Different approaches based on game theory and optimization-based techniques were employed. Extensive simulations were performed, for which the proposed framework achieved enhanced secrecy performance while maintaining low complexity, compared to
greedy association and positioning benchmarks.

\bibliographystyle{IEEEtran}
\bibliography{IEEEabrv,readme.bib}


\vfill

\end{document}